\documentclass[sn-basic]{sn-jnl}% Basic Springer Nature Reference Style/Chemistry Reference Style
\usepackage{amsmath}
\jyear{2022}%

\begin{document}

\title{A lipid-structured model of atherosclerosis with macrophage proliferation}


\author[1,2]{\fnm{Keith L.} \sur{Chambers}}\email{keith.chambers@maths.ox.ac.uk}

\author[1,3]{\fnm{Michael G.} \sur{Watson}}\email{michael.watson@sydney.edu.au}

\author*[1]{\fnm{Mary R.} \sur{Myerscough}}\email{mary.myerscough@sydney.edu.au}

\affil*[1]{\orgdiv{School of Mathematics and Statistics}, \orgname{The University of Sydney}, \orgaddress{ \city{Sydney}, \postcode{2006}, \state{NSW}, \country{Australia}}}

\affil[2]{\orgdiv{Mathematical Institute}, \orgname{The University of Oxford}, \orgaddress{\city{Oxford}, \postcode{OX2 6GG}, \state{Oxfordshire}, \country{United Kingdom}}}

\affil[3]{\orgdiv{School of Mathematics and Statistics}, \orgname{The University of New South Wales}, \orgaddress{ \city{Sydney}, \postcode{2052}, \state{NSW}, \country{Australia}}}

\abstract{We extend the lipid-structured model for atherosclerotic plaque development of \cite{ford2019lipid} to account for macrophage proliferation. Proliferation is modelled as a non-local decrease in the lipid structural variable that is similar to the treatment of cell division in size-structured models (e.g. \cite{efendiev2018functional}). Steady state analysis indicates that proliferation assists in reducing eventual necrotic core size and acts to spread the lipid load of the macrophage population amongst the cells. The relative contribution of plaque macrophages by proliferation and recruitment from the bloodstream is also examined. The model suggests that a more proliferative plaque differs from an equivalent (same lipid content and cell count) recruitment-dominant plaque only in the way lipid is distributed amongst the macrophages.}

\keywords{Proliferation, atherosclerosis, non-local, lipid- structured}



\maketitle

\section{Introduction}\label{sec:Intro}

Atherosclerosis is a chronic inflammatory disease of the artery wall \cite{Back_etal_2019,Wolf_Ley_2019}. The disease begins when disturbed blood flow creates a small lesion in the artery wall, allowing fatty compounds called lipids to enter from the bloodstream attached to low-density-lipoprotein (LDL) particles. The accumulation of LDL triggers an immune response which attracts circulating monocytes into the lesion that rapidly differentiate into macrophages upon entry. Macrophages consume the LDL lipid in a process called \textit{phagocytosis} and contribute to the further recruitment of monocyte-derived macrophages via inflammatory signalling \cite{Moore_etal_2013,Xu_etal_2019}. Over time the lesion may develop into an atherosclerotic plaque containing a large necrotic core of extracellular lipids, sourced from the death of lipid-laden macrophages. The rupture of such a plaque releases the necrotic core content into the bloodstream, where it can clot the blood and induce myocardial infarction or stroke. The mechanisms of early atherosclerosis that lead to necrotic core formation remain an active research topic \cite{Gonzalez_etal_2017}.

Atherosclerotic plaque development is driven by the dynamic interaction between macrophages and lipids, in addition to the relative rates at which these constituents enter and leave the plaque \cite{Moore_etal_2013}. Macrophages ingest lipid by consuming nearby LDL particles, apoptotic (dying) cells and necrotic material. They can play a protective role by removing lipid from the plaque, either by offloading lipid to high-density lipoprotein (HDL) particles for transport out of the artery wall, or by emigrating from the plaque. However, if a macrophage undergoes apoptosis (programmed cell death) and is not efficiently cleared by a live macrophage (a process called \textit{efferocytosis}), then the apoptotic cell will become secondarily necrotic. Necrotic cell death causes the internal lipid of the macrophage to spill into the extracellular environment. This is how the necrotic core grows. These processes suggest that the number of plaque macrophages and the lipid ingested by these cells are likely to be influential factors in the determination of plaque fate. An additional, and often under-appreciated, mechanism that may substantially alter these quantities is macrophage proliferation. This process is the focus of our modelling.

A significant number of plaque macrophages derive from the proliferation of macrophages that are already in the plaque, as opposed to the recruited monocyte-derived macrophages \cite{lhotak2016characterization, robbins2013local,Swirski_etal_2014, Takahashi_etal_2002}.  However, the role of macrophage proliferation in plaque development is not well understood. Proliferation may play a protective role by increasing the overall number of plaque macrophages and hence increasing the number of cells that can ingest  necrotic core material. However, proliferation may also serve as a significant source of plaque lipid. Scaglia \textit{et al.}  showed that an intracellular synthesis of lipids (fatty acids in particular) is required to complete cell-division \cite{Scaglia_etal_2014}. These lipids are likely used to form the membrane of the daughter cells \cite{Blank_etal_2017} and may contribute to the necrotic core upon secondary necrosis. Overall, it is not clear if macrophage proliferation is a net-protective effect \cite{Kim_etal_2021,Xu_etal_2019}. Hence it is not known whether therapies should promote or inhibit macrophage proliferation to reduce necrotic core growth. Mathematical modelling provides a means to explore this question.

Mathematical models of atherosclerosis that account for macrophage local proliferation are scarce. To the best of our knowledge, there are only two examples:  a stochastic model by Simonetto {\it et al.\ }\cite{Simonetto_etal_2017} and a spatio-temporal PDE model by Mukherjee {\it et al.\ }\cite{Mukherjee_etal_2020}. Both models follow the established convention of partitioning the macrophage population into ``macrophages" and ``foam cells" where foam cells are lipid-laden macrophages that take on a foamy appearance under the microscope \cite{Avgerinos_Neofytou_2019,Calvez_etal_2009,Chalmers_etal_2017,Chalmers_etal_2015}. We argue that macrophage local proliferation cannot be faithfully represented in such a framework. The crux of the issue is what happens when foam cells divide when they proliferate. Upon division, the internalised lipid of a foam cell is divided amongst the two daughter cells, but it is not clear how these daughter cells should be classed. If the foam cell is heavily lipid-laden, then the two daughter cells may contain enough lipid to both be classified as foam cells too. But if the foam cell contains only just enough lipid to meet the criterion to be classified as a foam cell, then the daughter cells will be classed as regular macrophages. The two models with proliferation mentioned above \cite{Mukherjee_etal_2020,Simonetto_etal_2017} avoid this ambiguity by assuming that foam cells do not divide. However, this assumption is at odds with experimental evidence which indicates that macrophage local proliferation occurs at every level of lipid accumulation at least in mice \cite{kim2018transcriptome}. A natural solution to the problem of foam cell division is to track the lipid content of the macrophages as a structural variable.

We use the model of \cite{ford2019lipid} for macrophage populations in atherosclerotic plaques as the foundation of this study on macrophage proliferation. The Ford model includes the lipid that macrophages contain as a structural variable. This includes lipid from ingested LDL, from apoptotic cells and endogenous lipid in the macrophages' membranes. The Ford model is a system of partial integro-differential equations and aimed to capture how the distribution of lipid in the live and apoptotic macrophage populations contributes to the formation of a necrotic core. Monocyte recruitment is included via a boundary condition, onloading and offloading of lipid by continuous advection, efferocytosis via a non-local convolution integral, and apoptosis and emigration via kinetic terms. Steady state analysis indicates that there are two qualitatively distinct profiles of the macrophage lipid distribution. The results demonstrate the important role of emigration and efferocytosis in reducing eventual necrotic core size. In this paper we extend the Ford model by adding proliferation. Mathematically, this introduces non-local terms to the set of integro-differential equations as well as, biologically, describing a means of reducing macrophage lipid loads.

The remainder of this paper is structured as follows. Section \ref{sec:Extended Ford model} describes the extended model which accounts for macrophage proliferation. Proliferation is modelled using a pantograph-style source term and local sink term. Similar terms are found in cell-division studies for size-structured  \cite{efendiev2018functional} and mass-structured models \cite{Sinko_Streifer_1971}. Section \ref{sec: results} contains the results of our model analysis. This includes a numerical simulation of time-dependent solutions and an analytical steady state analysis. Finally, we discuss the implications of our results in Section \ref{sec: discussion}. 

\section{Methods}\label{sec:Extended Ford model}

\subsection{Definitions}

Let $m(a,t)$ and $p(a,t)$ be the number density of the live and apoptotic macrophages in the lesion respectively. These densities are measured with respect to lipid content $a \geq a_0$ and depend on time $t \geq 0$. The quantity $a_0$ denotes the endogenous lipid content inherent to the macrophage membrane structure. The lipid content of the necrotic core is denoted by $N(t)$. The total number of live and apoptotic macrophages in the lesion, denoted $M(t)$ and $P(t)$ respectively, are found by integrating over all possible lipid contents:
\begin{align}
    M(t) &:= \int_{a_0}^{\infty} m(a,t) da, & P(t) &:= \int_{a_0}^{\infty} p(a,t) da. \label{eqn: MP defs}
\end{align}
Similarly, the total lipid content stored in the live and apoptotic macrophage populations are given by:
\begin{align}
    A_M(t) &:= \int_{a_0}^{\infty} a m(a,t) da, & A_P(t) &:= \int_{a_0}^{\infty} a p(a,t) da. \label{eqn: A_MA_Pdefs}
\end{align}

\subsection{Accounting for macrophage local proliferation}

Our treatment of macrophage local proliferation is predicated on the following assumptions:
\begin{enumerate}
    \item Macrophages divide at a constant rate, $\rho$, that is independent of the lipid content of the parent cell;
    \item Prior to division, the parent cell synthesises the $a_0$ endogenous lipid required to form the daughter cell membranes;
    \item Lipid synthesis and cell division occur on a faster timescale than macrophage population dynamics;
    \item Division produces two daughter cells that contain an equal amount of lipid.
\end{enumerate}
Assumption 1 is made primarily for simplicity. Lipid-dependent rates are to be considered in a future study (Watson \textit{et al.} 2022, in preparation). Assumption 2 accounts for experimental observations that \textit{de novo} lipid synthesis is required at the mitotic exit to form the membrane (e.g. the nuclear envelope lipid bilayer) required for the structural integrity of the daughter cells (Scaglia \textit{et al.} 2014, Sawicki \textit{et al.} 2019). The exclusion of lipid synthesis in our model would result in macrophages of lipid content $a < 2a_0$ being (unphysically) unable to divide. Assumption 3 is reasonable because mitosis typically lasts for 1 hour (Araujo \textit{et al.} 2016), whereas atherosclerosis progresses over months or years (Shah \textit{et al.} 2015). Finally, assumption 4 is made primarily for simplicity but is expected to be a good approximation when averaging over many division events. Indeed, preferential sorting of phagocytic cargo into one daughter cell was only observed for infectious cargo in the study by Luo \textit{et al.} (2008). 

\subsection{Model statement}
% Modelling assumptions covered in 2 parts: (1) = Ford 2019 assumptions, (2) = local proliferation

% The present study extends the model of Ford \textit{et al.} (2019), to which the interested reader is directed for further explanation. 

We consider the following extension to the lipid-structured model of \cite{ford2019lipid}:
\begin{align}
\begin{split}
    \frac{\partial m}{\partial t} + \Big( \frac{\lambda}{M} + \theta N \Big) \frac{\partial m}{\partial a} &= \eta \int_{a_0}^{a-a_0} m(a-a', t) p(a', t) da' - \eta P m  \\
    &\quad - (\beta + \gamma ) m + \underbrace{4\rho m( 2a - a_0, t) - \rho m}_{\text{local proliferation terms}},  \label{eqn: mdim}
\end{split} \\
    \frac{\partial p}{\partial t} &= \beta m - (\eta M + \nu) p, \label{eqn: pdim}\\
    \frac{dN}{dt} &=  \nu A_P - \theta M N, \label{eqn: Ndim}
\end{align}
where $a > a_0$ and $t > 0$. The above equations are coupled to the following boundary condition:
\begin{align}
    \Big( \frac{\lambda}{M} + \theta N \Big) m(a_0, t) &=  \alpha \frac{A_M - a_0 M}{\kappa + A_M - a_0 M}, & t &> 0. \label{eqn: bconddim}
\end{align}

\begin{figure}
    \centering
    \includegraphics[width = 1.0 \textwidth]{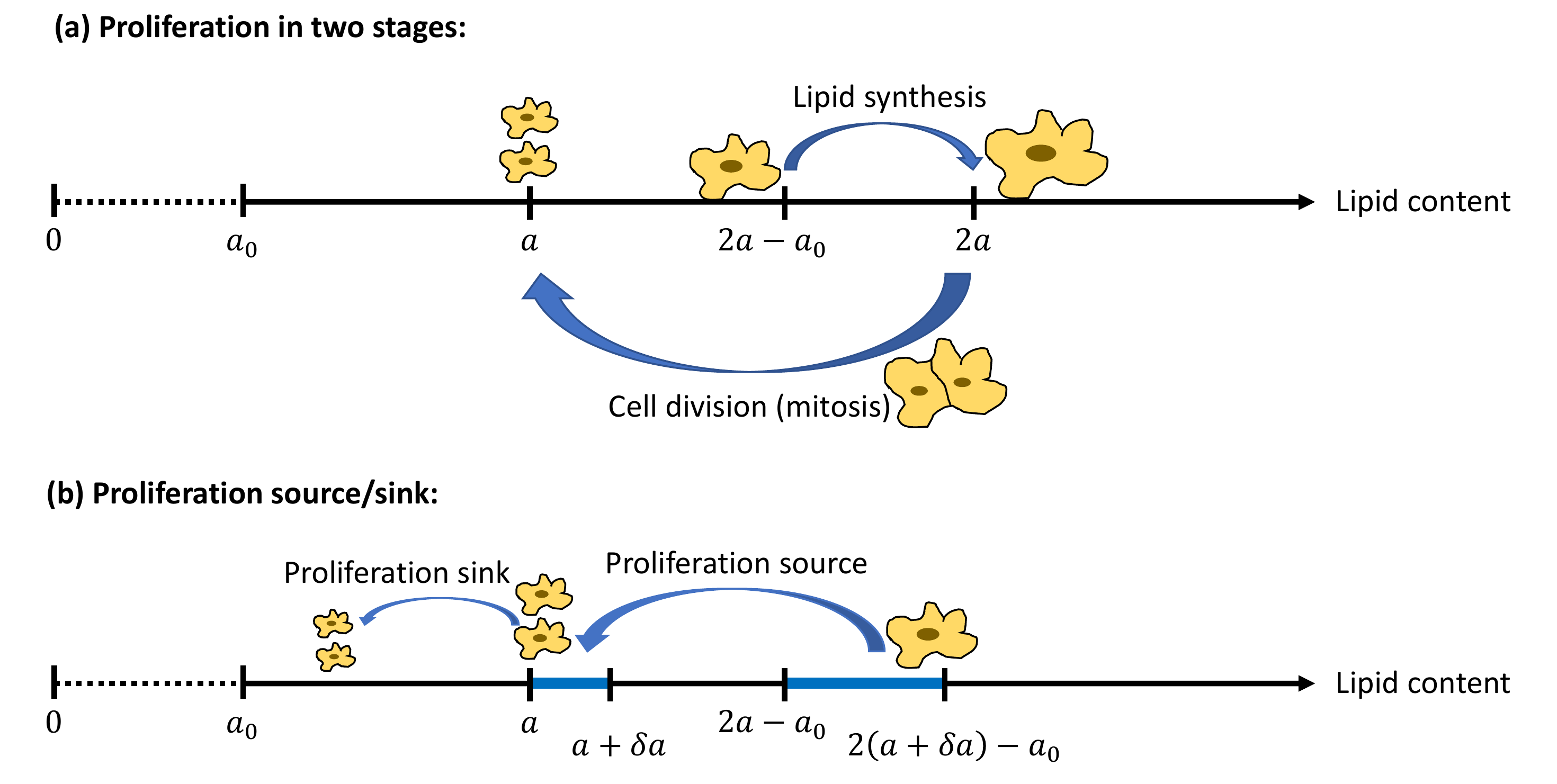}
    \caption{(a) Macrophage proliferation viewed as two stages. During proliferation, lipid synthesis increases the lipid content of the cell by $a_0$. Synthesis is followed by cell division, which splits the total lipid equally into the two daughter cells. The net result is that macrophages of lipid content $2a - a_0$ proliferate into two cells of lipid content $a$. Both processes occur on a faster timescale than the population dynamics which we model. (b) The source and sink processes accounted for in equation \eqref{eqn: mdim}.}
    \label{fig: proliferation diagram}
\end{figure}

Macrophage local proliferation is treated in the final two terms of equation \eqref{eqn: mdim}. The source term, $4 \rho m (2a - a_0, t)$, accounts for macrophages of lipid content $2a - a_0$ proliferating into two macrophages of lipid content $a$. The factor $4 = 2^2$ is due to each division event producing two daughter cells, and the fact that macrophages proliferating into the infinitesimal interval $(a, a + \delta a)$ are sourced from the interval $(2a - a_0, 2a - a_0 + 2 \delta a)$, which is twice as large. The sink term, $- \rho m(a, t)$, accounts for macrophages of lipid load $a$ dropping to a lower lipid content upon proliferation. The translation of the model assumptions into the source and sink terms described above is illustrated in Figure \ref{fig: proliferation diagram}.

The remaining terms are from the original model of Ford \textit{et al.} (2019), to which the interested reader is directed for detailed explanation. The advection term on the left side of equation \eqref{eqn: mdim} accounts for incremental changes in lipid content. This includes uptake of modLDL and offloading to HDL (net uptake rate $\propto \lambda$), and consumption of necrotic lipid (rate $\propto \theta$). By contrast, efferocytosis (rate $\propto \eta$) is modelled as a non-local increase in lipid content due to the consumption of whole apoptotic cells. The convolution source term in equation \eqref{eqn: mdim} accounts for all possible ways in which a macrophage can obtain a lipid content $a$ following the consumption of an apoptotic cell. It is to be interpreted as $0$ for $a < 2a_0$. The model also accounts for macrophage apoptosis (rate $\beta$), emigration from the lesion (rate $\gamma$) and the secondary necrosis of apoptotic cells (rate $\nu$). Finally, macrophage recruitment from the bloodstream is captured in the boundary condition \eqref{eqn: bconddim}. Recruited macrophages are assumed to carry only endogenous lipid, $a = a_0$. The rate of recruitment is an increasing function of the internalised lipid content of the live macrophage population, $A_M - a_0 M$. Maximal recruitment occurs at rate $\alpha$ and half-maximal recruitment occurs when $A_M - a_0 M = \kappa$. 

Additional ODEs for $M(t)$ and $P(t)$ can be derived by integrating equations \eqref{eqn: mdim} and \eqref{eqn: pdim} with respect to $a \in (a_0, \infty)$. The dynamics of $A_M(t)$ and $A_P(t)$ can be found similarly by multiplying by $a$ prior to integration. In this way, we obtain:
\begin{align}
    \frac{dM}{dt} &= \alpha \frac{A_M - a_0 M}{\kappa + A_M - a_0 M} + (\rho - \beta - \gamma) M, \label{eqn: Mdim}\\
    \frac{dP}{dt} &= \beta M - (\eta M + \nu) P, \label{eqn: Pdim}\\
    \frac{dA_M}{dt} &=  \alpha a_0 \frac{A_M - a_0 M}{\kappa + A_M - a_0 M} + \rho a_0 M -(\beta + \gamma)A_M \nonumber \\
    &\quad + \lambda + \theta MN + \eta M A_P, \label{eqn: A_Mdim}\\
    \frac{dA_P}{dt} &= \beta A_M - (\eta M + \nu) A_P. \label{eqn: A_Pdim}
\end{align}
The equations \eqref{eqn: Ndim}, \eqref{eqn: Mdim}-\eqref{eqn: A_Pdim} give a closed subsystem of ODEs for $M(t)$, $P(t)$, $A_M(t)$, $A_P(t)$ and $N(t)$ that can be solved independently of the non-local PDE system \eqref{eqn: mdim}-\eqref{eqn: bconddim}. We note that proliferation contributes to the increase in $M(t)$, due to division, and $A_M(t)$, due to lipid synthesis.

\subsubsection{Initial conditions}

Appropriate initial conditions must be prescribed in order to close the system of equations \eqref{eqn: mdim}-\eqref{eqn: A_Pdim}. Generically, we set:
\begin{align}
\begin{split}
        m(a,0) = m_0(a), &\quad p(a,0) = p_0(a), \\
    M(0) = M_0, \, \,  P(0) = P_0, \, \, A_M(0) &= A_{M0}, \, \, A_P(0) = A_{P0}, \, \, N(0) = N_0.
\end{split}
\end{align}
Here $M_0$, $P_0$, $A_{M0}$, $A_{P0}$ and $N_0$ are constants. The initial lipid distributions, $m_0(a)$ and $p_0(a)$, are assumed to be half-normal:
\begin{align}
    \frac{m_0(a)}{M_0} = \frac{p_0(a)}{P_0} = \frac{1}{a_\sigma} \sqrt{\frac{2}{\pi}} \exp \Big[ \frac{-(a-a_0)^2}{2 a_\sigma^2} \Big], \label{eqn: m0a def}
\end{align}
with a common variance determined by the constant $a_\sigma > 0$. It follows that $A_{M0} = \int_{a_0}^\infty a m_0(a) da$ and $A_{P0} =  \int_{a_0}^\infty a p_0(a) da$ satisfy the following relations:
\begin{align}
    A_{M0} &= M_0 \Big( a_0 + \sqrt{\frac{2}{\pi}} a_\sigma \Big), & A_{P0} &= P_0 \Big( a_0 + \sqrt{\frac{2}{\pi}} a_\sigma \Big),
\end{align}
from which we derive the equation:
\begin{align}
    M_0 =  \frac{\kappa \lambda \sqrt{2\pi}}{a_\sigma (\alpha \sqrt{2 \pi} a_\sigma - 2\lambda)},
\end{align}
by ensuring consistency with the boundary condition \eqref{eqn: bconddim}.

We note that biologically realistic initial conditions must satisfy $a_\sigma > \frac{\sqrt{2} \lambda}{\sqrt{\pi} \alpha}$ so that $M_0 > 0$. We further assume that the lesion is initially devoid of necrotic lipid: $N_0 = 0$,  and contains fewer apoptotic cells than live cells: $P_0 < M_0$. To satisfy this last condition, we arbitrarily set $P_0 = 0.5 M_0$.

\subsection{Non-dimensionalisation} \label{sec: nondim}

We scale macrophage lipid content in units of endogenous lipid, $a_0$, and time in units of mean macrophage lifespan, $\beta^{-1}$, by setting:
\begin{align}
    \tilde{a} &:= a_0^{-1}a, & \tilde{t} &:= \beta t. \label{eqn: a t scaling}
\end{align}
The remaining variables are non-dimensionalised as follows:
\begin{align}
    &\tilde{m} (\tilde{a}, \tilde{t}) := \frac{a_0 m(a,t)}{M(t)}, & &\tilde{p} (\tilde{a}, \tilde{t}) := \frac{a_0 p(a,t)}{P(t)}, \nonumber \\
    &\tilde{M}(\tilde{t}) := \frac{\beta}{\alpha} M(t), & &\tilde{P}(\tilde{t}) := \frac{\beta}{\alpha} P(t), \label{eqn: nondim} \\
    &\tilde{A}_M(\tilde{t}) := \frac{\beta}{a_0 \alpha} A_M(t), & &\tilde{A}_P(\tilde{t}) := \frac{\beta}{a_0 \alpha} A_P(t), & &\tilde{N}(\tilde{t}) := \frac{\beta}{a_0 \alpha} N(t). \nonumber
\end{align}

The scaling \eqref{eqn: nondim} ensures that 
\begin{equation}
\int_1^\infty \tilde{m} (\tilde{a}, \tilde{t}) d\tilde{a} = \int_1^\infty \tilde{p} (\tilde{a}, \tilde{t}) d\tilde{a} = 1, \label{eqn: normalised}
\end{equation}
and so $\tilde{m} (\tilde{a}, \tilde{t})$ and $\tilde{p} (\tilde{a}, \tilde{t})$ can be interpreted as probability distributions for the lipid contained in the live and apoptotic macrophage populations respectively. We use the same scale factor, $\beta / \alpha$, for the populations $M(t)$ and $P(t)$, and another common scale factor, $\beta / a_0 \alpha$, for the lipids $A_M(t), A_P(t), $ and $N(t)$. This is a different nondimensionalisation to the one used in \cite{ford2019lipid} and is chosen to facilitate comparison between the variables during analysis.

We also define a number of dimensionless parameters. These are listed in Table \ref{tab:parameters}. The first three parameters, $\tilde{\rho}$, $\tilde{\gamma}$ and $\tilde{\nu}$ pertain to cellular kinetics. We note that the dimensionless proliferation rate, $\tilde{\rho}$, plays a central role in the present study. The next three parameters, $\tilde{\eta}$, $\tilde{\theta}$ and $\tilde{\lambda}$, can be interpreted as dimensionless rates of lipid uptake. The parameter $\tilde{\lambda}$ differs in form from $\tilde{\eta}$ and $\tilde{\theta}$, ultimately due to the distinct treatment of LDL/HDL uptake from apoptotic/necrotic lipid uptake in \cite{ford2019lipid}. The constant $\tilde{\kappa}$ is the (dimensionless) accumulated live lipid required for half-maximal macrophage recruitment, linking lipid uptake to the cellular kinetics. Finally, the scale factor $\tilde{a}_\sigma$ determines the spread of the initial distributions $\tilde{m}(\tilde{a}, 0) = \tilde{p}(\tilde{a},0)$.

\begin{table}[]
\begin{tabular}{|c|c|l|}
\hline
\textbf{Parameter}     & \textbf{Definition}   & \textbf{Interpretation} \\ \hline
$\tilde{\rho}$                       &  $\frac{\rho}{\beta}$                     &  Dimensionless proliferation rate                   \\ \hline
$\tilde{\gamma}$                       &  $\frac{\gamma}{\beta}$                     &  Dimensionless emigration rate                       \\ \hline
$\tilde\nu$                       &   $\frac{\nu}{\beta}$                    & Dimensionless secondary necrosis rate                        \\ \hline
$\tilde\eta$                       &$\frac{\alpha \eta}{\beta^2}$                       & \begin{tabular}{@{}l@{}} Dimensionless uptake rate of apoptotic lipid \\ (efferocytosis) \end{tabular}                          \\ \hline
$\tilde{\theta}$                       &    $\frac{\alpha \theta}{\beta^2}$                       & Dimensionless uptake rate of necrotic lipid                         \\ \hline
$\tilde{\lambda}$                       & $\frac{\lambda}{a_0 \alpha}$                      & Dimensionless net rate of lipid uptake via LDL/HDL                        \\ \hline
$\tilde{\kappa}$                       &    $\frac{\kappa \beta}{a_0 \alpha}$                   &   \begin{tabular}{@{}l@{}} Dimensionless live cell accumulated lipid content for \\ half-maximal recruitment\end{tabular}                         \\ \hline 
$\tilde{a}_\sigma$                       &    $\frac{a_\sigma \beta}{a_0}$                   &   \begin{tabular}{@{}l@{}} Scale parameter that determines the spread of $\tilde{m}(\tilde{a}, 0).$\end{tabular} \\  \hline
\end{tabular}
\caption{Non-dimensional parameters that appear in equations \eqref{eqn: m nondim}-\eqref{eqn: init nondim}.}
\label{tab:parameters}
\end{table}

By applying the scaling \eqref{eqn: nondim} and definitions of Table \ref{tab:parameters} to the model \eqref{eqn: mdim}-\eqref{eqn: A_Pdim}, and dropping the tildes for notational convenience, we obtain the following dimensionless PDEs:
\begin{align}
\begin{split} \label{eqn: m nondim}
    \frac{\partial m }{\partial t} + \Big( \frac{\lambda}{M} + \theta N \Big) \frac{\partial m}{\partial a} &= \eta P \Big[ \int_1^{a-1} m(a-a',t)p(a',t) da' - m(a,t) \Big]  \\
    &\quad +\rho \big[ 4 m(2a-1,t) -  m(a,t) \big]  \\
    &\quad - \Big( \frac{1}{M} \frac{A_M - M}{\kappa + A_M - M} + \rho \Big) m(a,t),
\end{split} \\
\frac{\partial p}{\partial t} &= \frac{M}{P} \big[ m(a,t)-p(a,t) \big], \label{eqn: p nondim}
\end{align}
the boundary condition:
\begin{align}
    \Big( \frac{\lambda}{M} + \theta N \Big) m(1,t) &= \frac{1}{M} \frac{A_M - M}{\kappa + A_M - M}, \label{eqn: bcond nondim}
\end{align}
and the decoupled ODE system:
\begin{align}
    \frac{dM}{dt} &= \frac{A_M - M}{\kappa + A_M - M} + (\rho -1 -\gamma) M, \label{eqn: M nondim}\\
    \frac{dP}{dt} &= M - (\eta M + \nu) P, \label{eqn: P nondim}\\
\begin{split} \label{eqn: Am nondim}
    \frac{d A_M}{dt} &= \frac{A_M - M}{\kappa + A_M - M} + \rho M - (1 + \gamma) A_M  \\
    &\quad + \lambda + M (\eta A_P + \theta N),
\end{split} \\
    \frac{d A_P}{dt} &= A_M - (\eta M + \nu) A_P, \label{eqn: Ap nondim}\\
    \frac{dN}{dt} &= \nu A_P - \theta M N. \label{eqn: N nondim}
\end{align}
The model is closed with the initial conditions:
\begin{align}
\begin{split} \label{eqn: init nondim}
    &m(a,0) = p(a,0) = \frac{2}{a_\sigma \sqrt{2\pi}} \exp \Big( -\frac{(a-1)^2}{2 a_\sigma^2} \Big), \\
    &M(0) = 2P(0) = \frac{\kappa \lambda \sqrt{2 \pi}}{a_\sigma \big[ a_\sigma \sqrt{2 \pi} - 2\lambda \big]}, \\
    &A_M(0) = 2 A_P(0) = \frac{\kappa \lambda \sqrt{2 \pi}}{a_\sigma \big[ a_\sigma \sqrt{2 \pi} - 2\lambda \big]} \Big( 1 + \frac{2}{\sqrt{2\pi}} a_\sigma \Big), \\
    &N(0) = 0.
\end{split}
\end{align}

We note that time-dependent scaling of $m(a,t)$ and $p(a,t)$ in definitions \eqref{eqn: nondim} gives rise to substantial differences between equations \eqref{eqn: m nondim}, \eqref{eqn: p nondim} and their respective dimensional equivalents \eqref{eqn: mdim}, \eqref{eqn: pdim}. In equation \eqref{eqn: m nondim}, the terms of the final line do not have equivalent counterparts in equation \eqref{eqn: mdim} and act to enforce the normalisation condition \eqref{eqn: normalised}. In equation \eqref{eqn: p nondim}, $m(a,t)$ and $p(a,t)$ now appear on the right with a common factor so that that $p(a,t)$ evolves at a rate proportional to the difference $m(a,t) - p(a,t)$.

\subsection{Treatment of model parameters} \label{sec: parameter treatment}

The large parameter space (see Table \ref{tab:parameters}) poses a significant challenge to our analysis of the model \eqref{eqn: m nondim}-\eqref{eqn: init nondim}. Since the intention of the present study is to investigate the influence of proliferation on the model dynamics, we choose to fix the parameters that are peripheral to this purpose. Specifically, we take $\tilde{\nu} = 0.8$, $\tilde{\kappa} = 5$,  $\tilde{a}_\sigma = 2$ and $\tilde{\lambda} = 0.1$. 

The estimate for $\tilde{\nu} = \nu/\beta$ is based upon a dimensional apoptosis rate of $\beta = 0.05$ h$^{-1}$ and secondary necrosis rate of $\nu = 0.06$ h $^{-1}$. The stated apoptosis rate is the half-maximal value reported in the study \cite{thon2018quantitative}, in which the authors obtain estimates by fitting their ODE model to \textit{in vitro} data of lipid-laden macrophages. The secondary necrosis rate is derived from observations by \cite{collins1997major} that cell lysis occurs within 12-24h following apoptosis.

It is likely that $\tilde{\kappa}$ and $\tilde{a}_\sigma$ are both order 1 quantities since $a_0$ is a natural unit for macrophage lipid content. We assume further that $\tilde{\kappa} > \tilde{a}_\sigma$ since otherwise there would be significant macrophage recruitment even in the absence of LDL influx. To arbitrarily satisfy this requirement, we take $\tilde{\kappa} = 5$ and $\tilde{a}_\sigma = 2$. 

With $\tilde{\kappa}$ and $\tilde{a}_\sigma$ fixed, we see from equations \eqref{eqn: init nondim} that the choice of initial conditions amounts to a specification of $\tilde{\lambda}$. We take $\tilde{\lambda} = 0.1$ so that the initial macrophage population is small: $M(0) \approx 0.13$. This is to model the presence of tissue-resident macrophages in the lesion prior to LDL infiltration at $t = 0$. 

The remaining parameters, $\tilde{\rho}$, $\tilde{\gamma}$, $\tilde{\theta}$ and $\tilde{\eta}$ are considered over a range of values in our analysis. Although accurate estimation of these parameters is currently unfeasible due to a lack of quantitative \textit{in vivo} data, there are observations that inform our choice of these parameter values. Firstly, we assume that $\tilde{\eta} > \tilde{\theta}$ to be consistent with reports that the uptake of apoptotic cells is more efficient than uptake of necrotic material \cite{kojima2017role}. We note also that the dimensionless emigration rate is likely to take values near $\tilde{\gamma} \approx 0.25$. This is based on a dimensional emigration rate of $\gamma = 0.013$ h$^{-1}$, which is an intermediate value between the 12.6\% transmigration rate reported in \cite{angelovich2017quantification} and 20h residence time reported in \cite{ghattas2013monocytes}. Finally, we will assume throughout that $\tilde{\rho} < 1 + \tilde{\gamma}$. This is a restriction based upon equation \eqref{eqn: M nondim}, which predicts unbounded growth in $M(t)$ if $\tilde{\rho} \geq 1 + \tilde{\gamma}$. Biologically, this unphysical regime corresponds to a scenario where the lesion macrophages proliferate faster than they leave the system via apoptosis or emigration.

\subsection{Numerical solution scheme}

Numerical solutions for the equations \eqref{eqn: m nondim}-\eqref{eqn: init nondim} are obtained by using the \textit{method of lines} and integrating the resulting ODE system with the Wolfram Mathematica routine \textit{NDSolve}. The semi-infinite $a$-domain ($a \geq 1$) is approximated with a large finite interval, $1 \leq a \leq a_\text{max}$, and discretised uniformly. In equation \eqref{eqn: m nondim}, we approximate the lipid derivative $\frac{\partial m}{\partial a}$ using the second order upwinding scheme and the integral term via the trapezoidal rule. Finally, we note that the proliferation source term $4 \rho m(2a-1,t)$ is ill-defined for $2a-1 > a_\text{max}$. We therefore omit this term when $a > \frac{a_\text{max}+1}{2}$ in our numerical scheme. This omission is justified for large $a_\text{max}$ since $m(a,t)$ is a probability function and must therefore satisfy: $m(a,t) \rightarrow 0$ as $a \rightarrow \infty$. For the scenarios considered in this paper, we find that this error is negligible provided that $a_\text{max} \gtrapprox 30$.

\section{Results} \label{sec: results}

\subsection{Time-dependent solutions} \label{sec: time-dep}

\begin{figure}
    \centering
    \includegraphics[width = 1.0\textwidth]{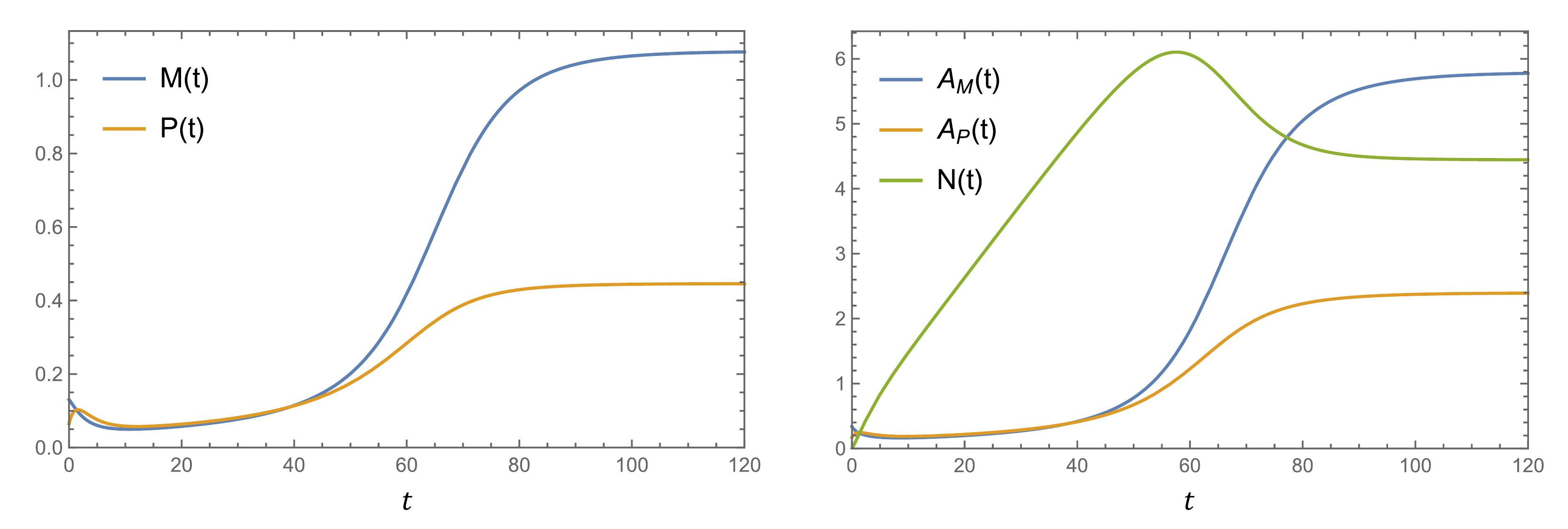}
    \caption{Time-dependent solutions for the ODE variables. The parameter values used are: $\rho = 0.8$, $\gamma = 0.25$, $\eta = 1.5$, $\theta = 0.4$.}
    \label{fig:ODE time dependent}
\end{figure}

Typical time-dependent solutions for the ODE variables are shown in Figure \ref{fig:ODE time dependent}. The dynamics appear to transition through four distinct stages. The first stage, spanning $0 \leq t \lessapprox 10$, is an initial transient in which $M(t)$ decreases slightly due to insubstantial recruitment relative to loss via apoptosis and emigration. There is correspondingly an initial increase in $P(t)$, followed by the function peaking and decreasing due to secondary necrosis. The lipid quantities $A_M(t)$ and $A_P(t)$ exhibit similar behaviour to $M(t)$ and $P(t)$ respectively. There is near-linear growth in $N(t)$ as the uptake rate of necrotic lipid (proportional to $\theta M(t)$) is small. For $10 \lessapprox t \lessapprox 50$, the system exhibits a slow increase in $M(t)$, $P(t)$, $A_M(t)$ and $A_P(t)$. These trends reflect a macrophage population that is gradually accumulating lipid from LDL uptake (modelled with $\lambda > 0$) and endogenous lipid from efferocytosis. The necrotic core continues to grow in an approximately linear fashion, albeit with a slightly decreased gradient. The dynamics enter a third stage when $N(t)$ grows large enough that uptake of necrotic lipid becomes comparable to uptake of apoptotic lipid: $\theta N(t) \sim \eta A_P(t)$. This occurs at $t \approx 50$ for the parameter values used in Figure \ref{fig:ODE time dependent}. The necrotic core becoming an additional substantial source of lipid uptake gives rise to a large increase in live lipid content, $A_M(t) - M(t)$. This leads to an increased recruitment of live macrophages, $M(t)$. The growth of $N(t)$ slows before the function peaks and begins decreasing as $t$ increases. Finally, the system reaches an equilibrium at $t \approx 120$.

\begin{figure}
    \centering
    \includegraphics[width = 1.0\textwidth]{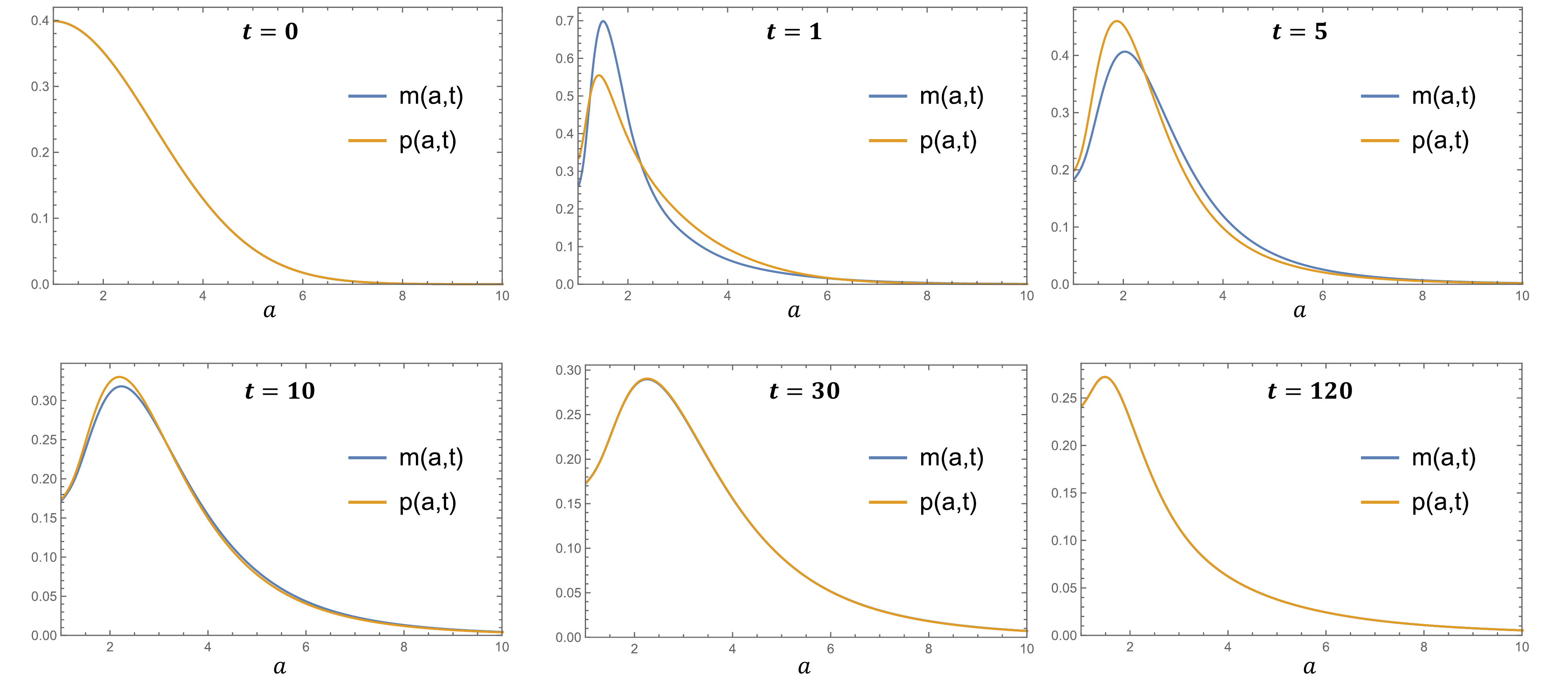}
    \caption{Time evolution  of the lipid distributions $m(a,t)$ and $p(a,t)$. The parameter values used are: $\rho = 0.8$, $\gamma = 0.25$, $\eta = 1.5$, $\theta = 0.4$.}
    \label{fig:PDE time dependent}
\end{figure}

The time evolution of the lipid distributions $m(a,t)$ and $p(a,t)$ is shown in Figure \ref{fig:PDE time dependent}. The simulation begins with $m(a,t) = p(a,t)$ as half-normal distributions, in accordance with the initial conditions \eqref{eqn: init nondim}. For very early times, $0 < t \lessapprox 1$, the live lipid distribution, $m(a,t)$, becomes increasingly concentrated towards lower lipid loads. Since $m(1,t)$ (which is proportional to the recruitment rate) decreases from its initial value, this early change is likely due to macrophage proliferation as opposed to recruitment. As $t$ increases between $1 \lessapprox t \lessapprox 30$, lipid uptake drives a gradual increase in the tail values of $m(a,t)$, reflecting a greater proportion of highly lipid-laden macrophages. As the system approaches equilibrium near $t = 120$, $m(a,t)$ becomes concentrated towards $a = 1$ once again. This final overall decrease in lipid load is likely driven by increased macrophage recruitment, rather than  proliferation, as can be seen by the increase in $m(1,t)$. We note that $p(a,t)$ is largely in agreement with $m(a,t)$ except at early but nonzero values of $t$. This is due to the form of equation \eqref{eqn: p nondim} which states that $p(a,t)$ evolves towards $m(a,t)$ at a rate proportional to $\frac{M(t)}{P(t)}$. Since $\frac{M(t)}{P(t)}$ is an order 1 quantity throughout the simulation (see Figure \ref{fig:ODE time dependent}), the rapid early change in $m(a,t)$ for $0 < t \lessapprox 1$ causes a disagreement between $m(a,t)$ and $p(a,t)$ early in the simulation. This disagreement is resolved on the order 1 timescale, with $m(a,t) \approx p(a,t)$ at $t = 10$.

Numerical simulations indicate that the system evolves towards a steady state, as indicated by Figures \eqref{fig:ODE time dependent} and \eqref{fig:PDE time dependent}. For the parameter values we considered, steady state is typically effectively attained within $t = 200$ macrophage lifetimes, corresponding to 167 days. Atherosclerotic plaques can exist for decades in humans, and so the steady state will give biologically useful information \cite{kusuma2018escalation}.

\subsection{Steady state analysis: ODE subsystem}

Let $M^\star$, $P^\star$, $A_M^\star$, $A_P^\star$ and $N^\star$ denote the steady values of the ODE variables. These constants can be found by setting the time derivatives in equations \eqref{eqn: M nondim}-\eqref{eqn: N nondim} to zero and solving the resulting system of five algebraic equations. Upon doing so, we find that each of the steady values can be expressed in terms of $M^\star$:
\begin{align}
\begin{aligned} \label{eqn: ODE steady}
    P^\star &= \frac{M^\star}{\eta M^\star + \nu}, & A_M^\star &= M^\star + \kappa \Big( \frac{1}{1- ( 1 +\gamma - \rho) M^\star} -1 \Big), \\
    A_P^\star &= \frac{A_M^\star}{\eta M^\star + \nu}, & N^\star &= \frac{\nu A_P^\star}{\theta M^\star},
\end{aligned}
\end{align}
and that $M^\star$ itself satisfies the quadratic equation:
\begin{align}
    (1 + \gamma - \rho) (M^\star)^2 + \big[ \lambda -1 + \gamma (\kappa + \gamma \kappa + \lambda) - \rho (\lambda + \gamma \kappa) \big] M^\star - \lambda = 0. \label{eqn: M quadratic}
\end{align}

Only one of the two possible solutions to equation \eqref{eqn: M quadratic} corresponds to a valid steady state. This can be seen by considering the product of the roots: $\frac{- \lambda}{ 1 + \gamma - \rho}$. Using that $\rho < 1 + \gamma$ (see Section \ref{sec: parameter treatment}), we deduce that the product of roots is negative. Since the coefficients of equation \eqref{eqn: M quadratic} are real, it follows that one solution is positive and the other negative. The candidate steady state is the positive solution:
\begin{align}
\begin{split}
    M^\star &= - \frac{\lambda -1 + \gamma (\kappa + \gamma \kappa + \lambda) - \rho (\lambda + \gamma \kappa)}{2(1+\gamma - \rho)} \\
    &\quad + \frac{\sqrt{\big[ \lambda -1 + \gamma (\kappa + \gamma \kappa + \lambda) - \rho (\lambda + \gamma \kappa) \big]^2 + 4 \lambda (1 + \gamma - \rho)}}{2(1+\gamma - \rho)}. \label{eqn: Mstar}
\end{split}
\end{align}
It can be shown by substituting the solution \eqref{eqn: Mstar} into the relations \eqref{eqn: ODE steady} that $A_M^\star > M^\star$, $A_P^\star > P^\star$ and $N^\star > 0$. Hence, a unique and valid steady state always exists for the model, and is given by equation \eqref{eqn: Mstar}. Although it is difficult to prove, numerical simulations of the ODE subsystem indicate that the steady state is indeed stable. 

\subsubsection{Influence of proliferation on the ODE steady state} \label{sec: ODE parameter dependence}

\begin{figure}
    \centering
    \includegraphics[width = 1.0\textwidth]{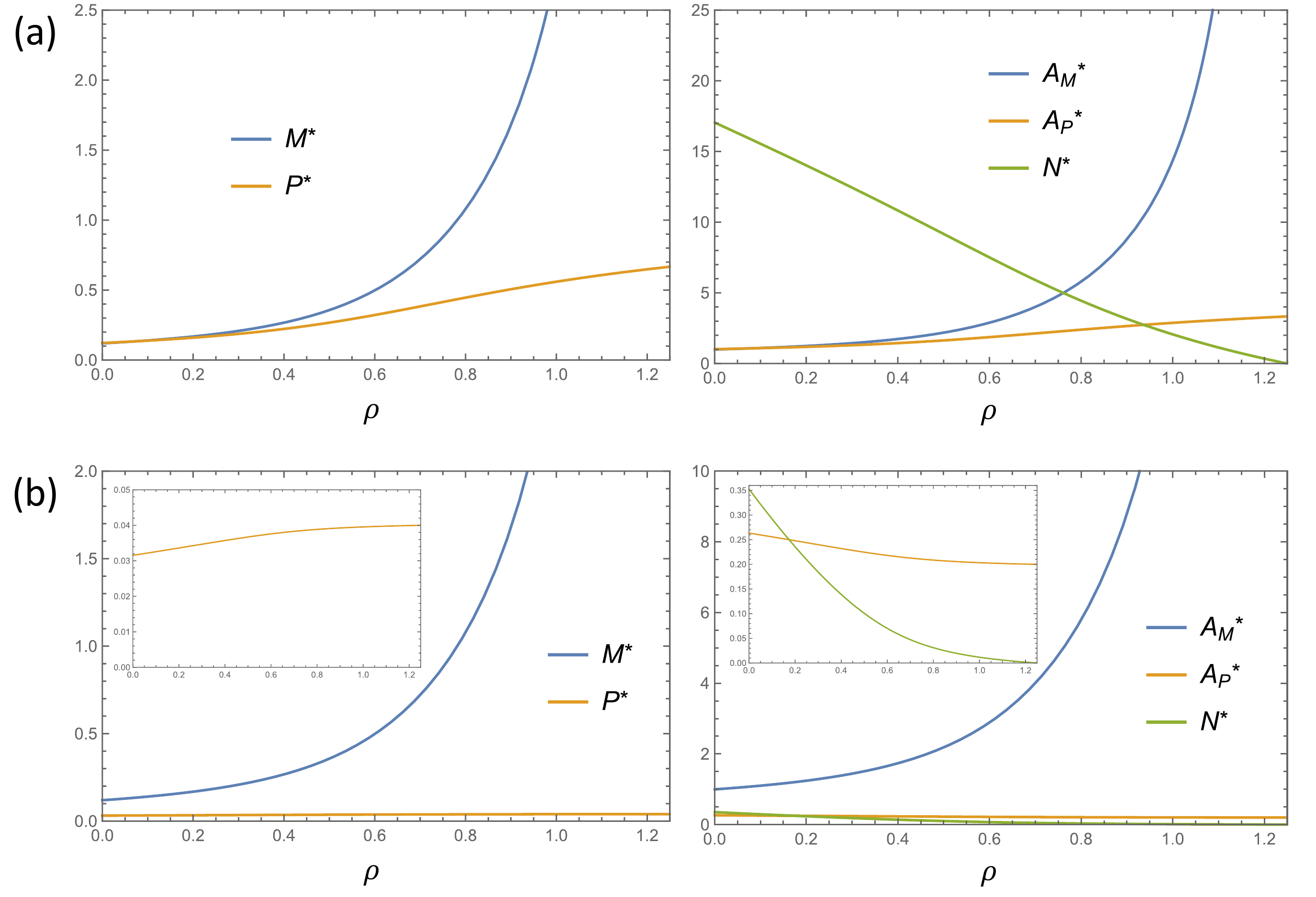}
    \caption{Dependence of the ODE equilibrium values on the proliferation rate, $\rho$. The model output for two sets of parameter values is shown, labelled (a) and (b). Set (a) uses $\gamma = 0.25$, $\eta = 1.5$, $\theta = 0.4$ so that $\eta < \frac{(1+\gamma) \nu}{\lambda}$. Set (b) uses higher cellular lipid uptake rates: $\gamma = 0.25$, $\eta = 25$, $\theta = 5$ so that $\eta > \frac{(1+\gamma) \nu}{\lambda}$. The two sets exhibit opposing trends for $A_P^\star$.}
    \label{fig:ODE equilibrium trends}
\end{figure}

The steady values of the ODE variables change monotonically as the proliferation parameter, $\rho$, is increased from 0 to $1 + \gamma$. We prove this result in Appendix \ref{secA1} by taking partial derivatives of the equilibrium values with respect to $\rho$. Explicitly, we find that $M^\star$, $P^\star$ and $A_M^\star$ are increasing functions of $\rho$, that $N^\star$ is a decreasing function of $\rho$, and that $A_P^\star$ can be either monotone increasing or decreasing with $\rho$ depending on whether $\eta < \frac{(1+\gamma) \nu}{\lambda}$ or $\eta > \frac{(1+\gamma) \nu}{\lambda}$ respectively. Moreover,
we can use equations \eqref{eqn: ODE steady} and \eqref{eqn: Mstar} to explicitly derive the limiting values of the ODE steady state variables. Explicitly, we find the following limits as $\rho \rightarrow (1+\gamma)^{-}$:
\begin{align}
    M^\star,  A_M^\star \rightarrow \infty, \quad P^\star \rightarrow \frac{1}{\eta}, \quad A_P^\star \rightarrow \frac{1+\gamma}{\gamma \eta}, \qquad \text{and } \quad N^\star \rightarrow 0. \label{eqn: ODE steady limits}
\end{align}
Hence, the model predicts that a large enough proliferation rate is always capable of eliminating the necrotic core, albeit at the expense of having an extremely large cell population in the lesion. Plots of the ODE equilibrium values against $\rho$ are given in Figure \ref{fig:ODE equilibrium trends}.

The result that $M^\star$ and $P^\star$ are increasing functions of $\rho$ is unsurprising since cell division directly increases the number of cells in the lesion. It is also expected that $A_M^\star$ increases with $\rho$ due to the lipid synthesis that occurs with proliferation. Interestingly, the results indicate that when the efferocytosis rate is small enough, $\eta < \frac{(1+\gamma) \nu}{\lambda}$, increases in the proliferation rate give rise to a smaller necrotic core, $N^\star$, despite increasing the amount of apoptotic lipid in the lesion, $A_P^\star$. In this case, the reduction in necrotic core size is therefore conclusively due to an increase in the consumption of necrotic lipid by the live macrophage population. When $\eta > \frac{(1+\gamma) \nu}{\lambda}$, increases to the proliferation rate also decrease eventual necrotic core size by reducing the amount of apoptotic lipid in the lesion (which sources the necrotic core via secondary necrosis). However, as indicated in Figure \ref{fig:ODE equilibrium trends}(b), the reduction in $A_P^\star$ is minimal in comparison to the increase in necrotic lipid uptake, which is proportional to $\theta M^\star$. Hence, in both cases the primary mechanism by which increases in the proliferation rate reduce eventual necrotic core size is by increasing the total rate of necrotic lipid consumption at equilibrium. 

\begin{figure}
    \centering
    \includegraphics[width = 1.0\textwidth]{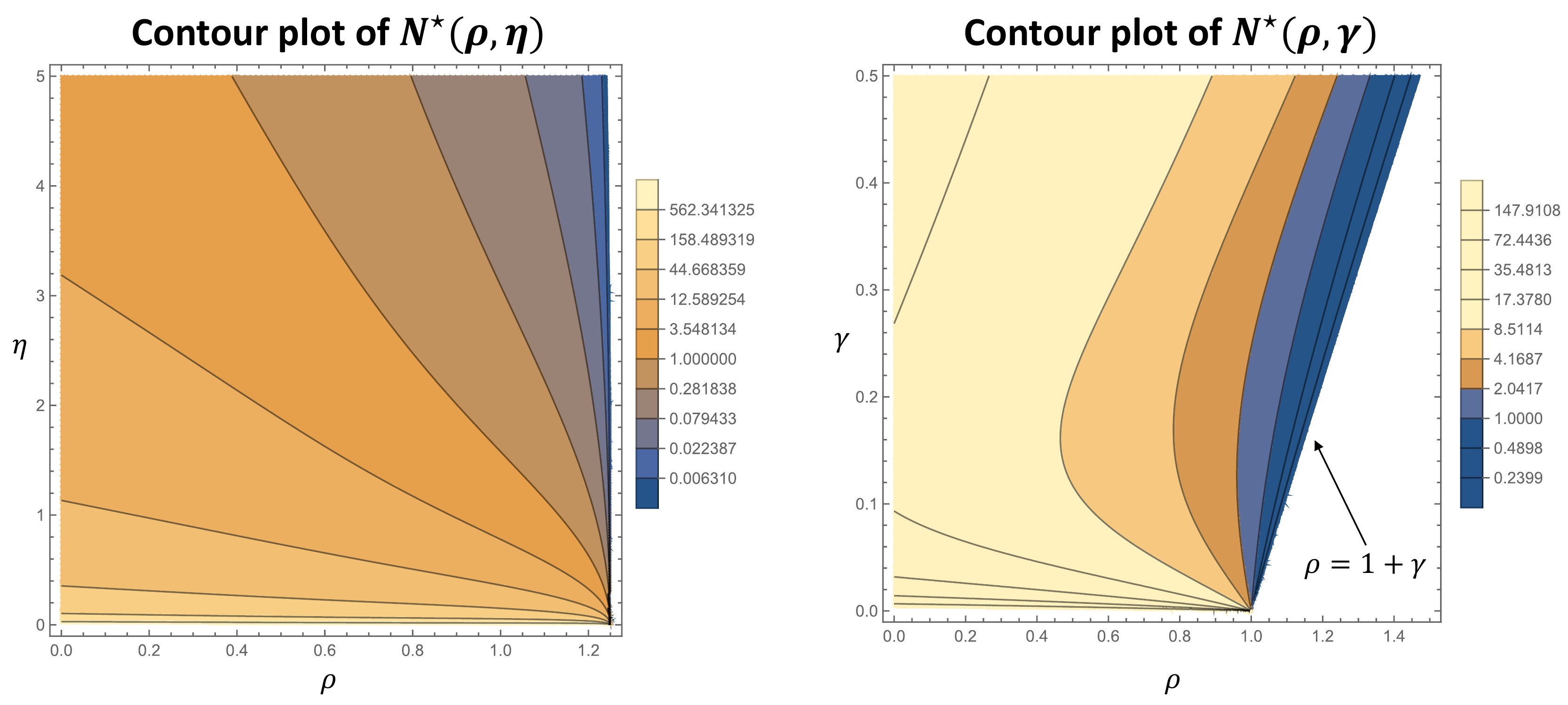}
    \caption{Contour plots of the equilibrium necrotic core size, $N^\star$. The left plot illustrates the dependence on $\rho$ and $\eta$, the rates of proliferation and efferocytosis respectively. We take $\theta = \frac{\eta}{2}$ to ensure that $\theta < \eta$ (see Section \ref{sec: ODE parameter dependence}) and set $\gamma = 0.25$. The right plot shows the dependence on $\rho$ and the emigration rate, $\gamma$. We use $\eta = 1.5$ and $\theta = 0.4$. The equilibrium only exists for $\rho < 1 + \gamma$, as indicated. Note that we use a logarithmic scale for the contours of both plots since $N^\star$ varies over a large range of values.}
    \label{fig: N contours}
\end{figure}

To better illustrate the role of proliferation in determining the equilibrium necrotic core size, $N^\star$, we vary $\rho$ with other model parameters simultaneously in Figure \ref{fig: N contours}. Consider firstly the left plot, in which $N^\star$ is varied as a function of $(\rho, \eta)$. Here we also take $\theta = \frac{\eta}{2}$ to arbitrarily satisfy the requirement that $\eta > \theta$ (see Section \ref{sec: ODE parameter dependence}). The plot indicates that increases to the proliferation rate most effectively reduce necrotic core size when $\eta$ is large (where the contours are more vertical). Indeed when $\eta$ is small the contours are approximately horizontal. Hence, increases in the proliferation rate are ineffective at decreasing necrotic core size when efferocytosis is defective. Consider now the right plot of Figure \ref{fig: N contours}, showing the dependence of $N^\star$ on $(\rho, \gamma)$. Interestingly, the plot shows a non-monotone dependence of $N^\star$ on $\gamma$ for fixed $\rho < 1$, indicating that there is a nonzero emigration rate ($\gamma \approx 0.16$ in the figure) for which $N^\star$ is minimised for given $\rho$. Furthermore, the plot indicates that $N^\star$ monotonically increases with $\gamma$ for $\rho > 1$. This result suggests, counter-intuitively, that increasing the migratory propensity of a self-replenishing lesion macrophage population will increase necrotic core size.

\subsection{Steady state analysis: $m(a,t)$ and $p(a,t)$}

We turn our focus to the steady state macrophage lipid distributions. Setting the time derivative to zero in equation \eqref{eqn: p nondim} gives $p = m$. Hence, the live and apoptotic macrophages have identical lipid distributions at steady state. It then follows from equations \eqref{eqn: m nondim} and \eqref{eqn: bcond nondim} that the equilibrium live macrophage lipid distribution, $m^\star(a) := \lim_{t \rightarrow \infty} m(a,t)$, satisfies the following boundary value problem:
\begin{align}
    \frac{d m^\star}{d a} &= k_1 \int_1^{a-1} m^\star (a-a') m^\star(a') da' - k_2 m^\star(a) + k_3 m^\star(2a-1), \label{eqn: mstar ode} \\
    m^\star(1) &= m^\star_1, 
\end{align}
where $k_1$, $k_2$, $k_3$ and $m^\star_1$ are constants defined in terms of the ODE equilibrium values:
\begin{align}
    k_1 &= \Big( \frac{\lambda}{M^\star} + \theta N^\star \Big)^{-1} \eta P^\star, & k_2 &= \Big( \frac{\lambda}{M^\star} + \theta N^\star \Big)^{-1} (\eta P^\star + 1 + \gamma + \rho), \\
    k_3 &= 4 \Big( \frac{\lambda}{M^\star} + \theta N^\star \Big)^{-1} \rho, & m^\star_1 &= \Big( \frac{\lambda}{M^\star} + \theta N^\star \Big)^{-1} \frac{1}{M^\star} \frac{A^\star_M - M^\star}{\kappa + A_M^\star - M^\star}.
\end{align}

The combined delay and advanced dependence on $a$ in equation \eqref{eqn: mstar ode} makes it difficult to find a general formula for $m^\star(a)$. The case $k_3 = 0$ (no proliferation) is solved in \cite{ford2019lipid} by partitioning the domain $a \geq 1$ into unit intervals and solving successively on each interval. An analytical solution can also be found for the case $k_1 = 0$ (no efferocytosis) using the Laplace transform. It is given by:
\begin{align}
    m^\star(a) = m^\star_1 \sum_{j = 0}^\infty \Big( \frac{k_3}{2k_2} \Big)^j \sum_{i = 0}^j \frac{2^i}{\prod_{\ell = 0, \ell \neq i}^j (1-2^{i-\ell})} e^{-2^i k_2 (a-1)}. \label{eqn: no use}
\end{align}
Our method, presented in Appendix \ref{secA2}, is inspired by Hall and Wake, who solved a similar problem in a model for cell growth \cite{hall1989functional}. Unfortunately the solution \eqref{eqn: no use} is too complicated to be biologically insightful and cannot, to the best of our knowledge, be manipulated into a closed-form expression. 

Although a closed-form solution to equation \eqref{eqn: mstar ode} is unlikely to exist in full generality, we show below that the standard summary statistics do have closed-form expressions.

\subsubsection{Summary statistics for $m^\star (a)$}

The statistical moments of $m^\star(a)$ are given by:
\begin{align}
    \varphi_n := \int_1^\infty a^n m^\star (a) da, \qquad n = 0, 1, 2, \dots .
\end{align}
Our non-dimensionalisation \eqref{eqn: nondim} ensures that $\varphi_0 = 1$ and $\varphi_1 = \frac{A_M^\star}{M^\star}$. Analytical expressions for the higher order moments can be found recursively by multiplying equation \eqref{eqn: mstar ode} by $a^n$ and integrating over $a \in (1, \infty)$. Doing so and solving for $\varphi_n$ gives the full-history recurrence:
\begin{align}
    \varphi_n = \frac{m^\star_1 + n \varphi_{n-1} + k_3/2^{n+1} + \sum_{j=1}^{n-1} \binom{n}{j} \varphi_j (k_1 \varphi_{n-j} + k_3/2^{n+1})}{k_2 - 2k_1 - k_3/2^{n+1}}, \label{eqn: recursion}
\end{align}
for each $n \geq 2$. Using the recursion \eqref{eqn: recursion}, closed-form expressions can be obtained for the standard summary statistics of a continuous probability distribution. For our analysis of $m^\star (a)$, we will use the mean, $\mu$, and skewness, $\tilde{\mu}_3$, defined by:
\begin{align}
    \mu &= \varphi_1, &  \tilde{\mu}_3 = \frac{\varphi_3 - 3 \varphi_1 (\varphi_2 - \varphi_1^2)-\varphi_1^3}{(\varphi_2 - \varphi_1^2)^{3/2}}. \label{eqn: summary stats}
\end{align}
Note that we do not explicitly consider the standard deviation of $m^\star(a)$ in the analysis below since the distribution is highly asymmetric. 

\subsubsection{Influence of proliferation on $m^\star (a)$}

\begin{figure}
    \centering
    \includegraphics[width = 1.0\textwidth]{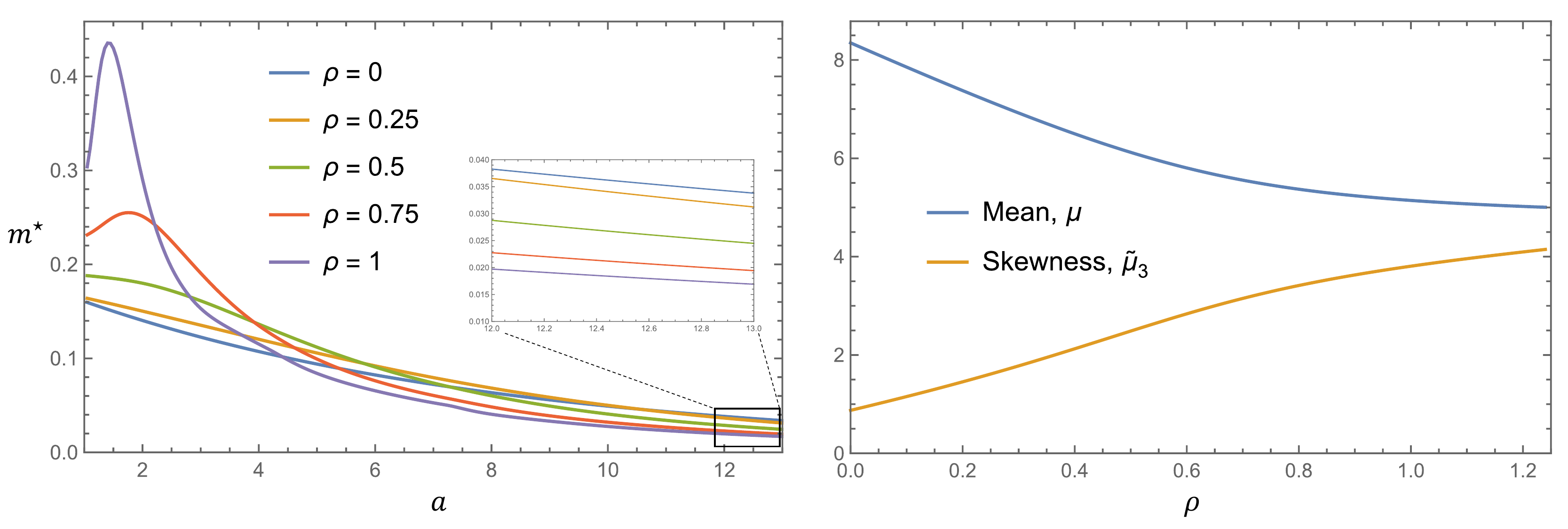}
    \caption{Influence of the proliferation rate, $\rho$, on the equilibrium lipid distribution, $m^\star (a)$. The left plot shows $m^\star(a)$ for various values of $\rho$. The right plot shows the corresponding summary statistics, computed using the formulae \eqref{eqn: summary stats}. The parameter values are set to: $\gamma = 0.25$, $\eta = 1.5$ and $\theta = 0.4$.}
    \label{fig:m(a) increasing rho}
\end{figure}

Plots of $m^\star (a)$ for various values of $\rho$ are shown in Figure \ref{fig:m(a) increasing rho}. These solutions were obtained by numerically simulating the time-dependent system \eqref{eqn: m nondim}-\eqref{eqn: init nondim} to $t = 200$. The mean and skewness are also plotted as functions of $\rho$. We find that increasing $\rho$ reduces the average lipid burden per macrophage, $\mu$, and increases the skewness, $\tilde{\mu}_3$. Correspondingly, as $\rho$ increases,
the tail of $m^\star (a)$ thins and the left end of the distribution grows. Interestingly, we see that a local maximum develops near $a = 1$ for $\rho$ large enough.

\begin{figure}
    \centering
    \includegraphics[width = 1.0\textwidth]{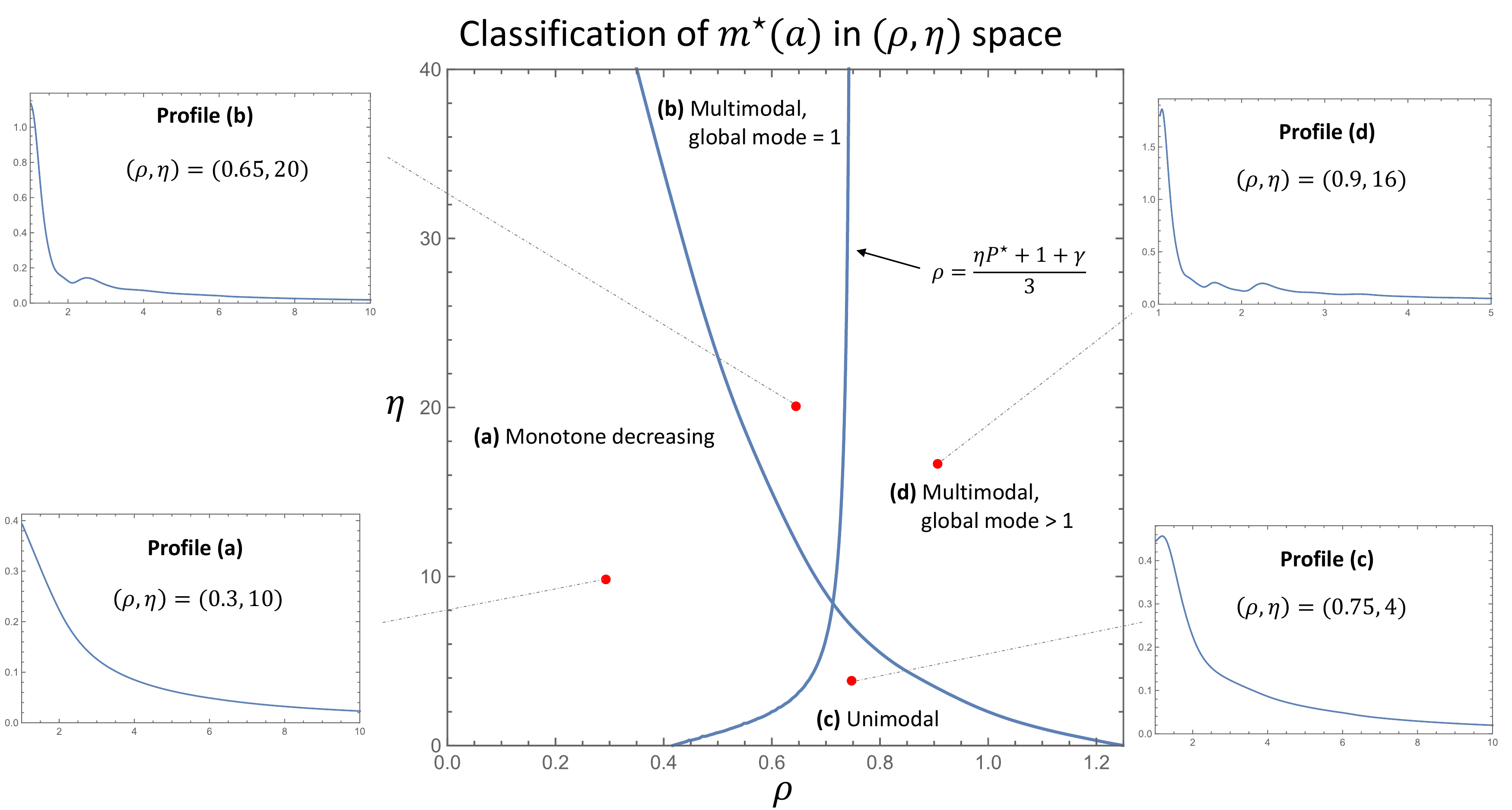}
    \caption{The four qualitatively distinct profiles of $m^\star (a)$ can be resolved in the $(\rho, \eta)$ subspace. In the central diagram, the curve separating regions (a) and (b) from (c) and (d) was is given by $\rho = \frac{\eta P^\star + 1 + \gamma}{3}$, and was plotted using the ODE equilibrium solutions \eqref{eqn: ODE steady} and \eqref{eqn: Mstar}. The curve separating regions (a) and (c) from (b) and (d) was found numerically by collecting 20 sample points and linking with second order interpolation. Examples for each profile are shown, using the $(\rho, \eta)$ values indicated by the red points. The remaining parameters are set to: $\gamma = 0.25$ and $\theta = 0.4$.}
    \label{fig: m(a) profile classification}
\end{figure}

The model exhibits four qualitatively distinct profiles for $m^\star (a)$. These can be resolved in the $(\rho, \eta)$ parameter-subspace, as seen in Figure \ref{fig: m(a) profile classification}. If $\rho$ is small enough (quantified below), then $m^\star (a)$ either decreases monotonically, profile (a), or contains a series of peaks that are spaced approximately an integer apart, profile (b). These profiles are also observed in \cite{ford2019lipid}. For $\rho$ large enough, the global maximum shifts from $a = 1$ to a local maximum at some $a_\text{max} > 1$. The precise condition that must be satisfied for this to occur is: $\rho > \frac{\eta P^\star + 1 + \gamma}{3}$, which is found by setting $\frac{d m^\star}{d a} \big\vert_{a = 1} > 0$ in equation \eqref{eqn: mstar ode}. When the above condition holds, small values of $\eta$ result in a unimodal profile for $m^\star (a)$, profile (c). As $\eta$ is increased, a smaller secondary peak appears at $2 a_\text{max}$ due to macrophages containing $a_\text{max}$ lipid consuming equally lipid-laden apoptotic cells, profile (d). As seen in the figure, a tertiary peak can also be seen in some cases at $(2 a_\text{max} + 1)/2$ due to the proliferation of macrophages in the secondary peak.

\subsection{Proliferation vs. monocyte recruitment} \label{sec: relative contribution}

Macrophages in early atherosclerosis derive from either bloodstream recruitment or local proliferation (that is, proliferation in the artery wall). In this subsection, we use our model to explore how the relative contribution of these two processes may impact the eventual constitution of the plaque. Since the recruitment parameter, $\alpha$, appears throughout our non-dimensionalisation \eqref{eqn: nondim}, we find it more convenient to work with dimensional quantities in the paragraphs below. These are adorned with a hat to avoid confusion.

We investigate the relative contribution of proliferation and recruitment by considering changes in $\hat{\rho}$ (the proliferation rate) and $\hat{\alpha}$ (proportional to the recruitment rate). As a control, we assume that $\hat{\rho}$ and $\hat{\alpha}$ are changed in such a way that the number of live macrophages in the plaque at steady state, $\hat{M}^\star$, remains fixed. All other parameters are also held constant. To derive the algebraic constraint the above conditions put on $\hat{\rho}$ and $\hat{\alpha}$, we set $\hat{M^\star}_{\hat{\alpha} = \hat{\alpha}_1, \hat{\rho} = \hat{\rho}_1} = \hat{M^\star}_{\hat{\alpha} = \hat{\alpha}_2, \hat{\rho} = \hat{\rho}_2}$ and expand using the dimensional form of equation \eqref{eqn: Mstar}. The equation reduces to 
\begin{equation}
    \frac{\hat{\beta} + \hat{\gamma} - \hat{\rho}_1}{\hat{\alpha}_1} = \frac{\hat{\beta} + \hat{\gamma} - \hat{\rho}_2}{\hat{\alpha}_2}, \label{eqn: alpha rho constraint}
\end{equation}
revealing that $\hat{\rho}$ and $\hat{\alpha}$ must be adjusted such that the ratio $\frac{\hat{\beta} + \hat{\gamma} - \hat{\rho}}{\hat{\alpha}}$ is fixed. 

When $\tilde{\rho}$ and $\tilde{\alpha}$ are changed according to the constraint \eqref{eqn: alpha rho constraint}, we find that there is no change in the total apoptotic population, $\hat{P}^\star$, live or apoptotic lipid totals, $\hat{A}_M^\star$, $\hat{A}_P^\star$, or necrotic core size, $\hat{N}^\star$. Hence, the model predicts that these quantities are independent of the relative contribution of proliferation and recruitment, in the sense described above. This result can be seen analytically by considering the dimensional versions of the steady state solutions \eqref{eqn: ODE steady}:
\begin{align}
\begin{aligned}
    \hat{P}^\star &= \frac{\hat{\beta} \hat{M}^\star}{\hat{\eta} \hat{M}^\star + \hat{\nu}}, & \hat{A}_M^\star &= \hat{a}_0 \hat{M}^\star + \hat{a}_0 \hat{\kappa} \frac{\big( 1 + \frac{\hat{\gamma}}{\hat{\beta}} - \frac{\hat{\rho}}{\hat{\beta}} \big) \hat{M}^\star}{1 - \frac{\hat{\beta} + \hat{\gamma} - \hat{\rho}}{\hat{\alpha}} \hat{M}^\star}, \\
    \hat{A}_P^\star &= \frac{\beta \hat{A}_M^\star}{ \hat{\eta} \hat{M}^\star + \hat{\nu}}, & \hat{N}^\star &= \frac{\hat{\nu} \hat{A}_P^\star}{ \hat{\theta} \hat{M}^\star}.
\end{aligned}
\end{align}
Since $\hat{M}^\star$ and $\frac{\hat{\beta} + \hat{\gamma} - \hat{\rho}}{\hat{\alpha}}$ are fixed quantities, it follows that $\hat{P}^\star$ and $A_M^\star$ are also fixed. The same can then be concluded for $\hat{A}_P^\star$ and $\hat{N}^\star$, which depend on $\tilde{A}_M^\star$ and $\tilde{A}_P^\star$ respectively.

\begin{figure}
    \centering
    \includegraphics[width = 1.0\textwidth]{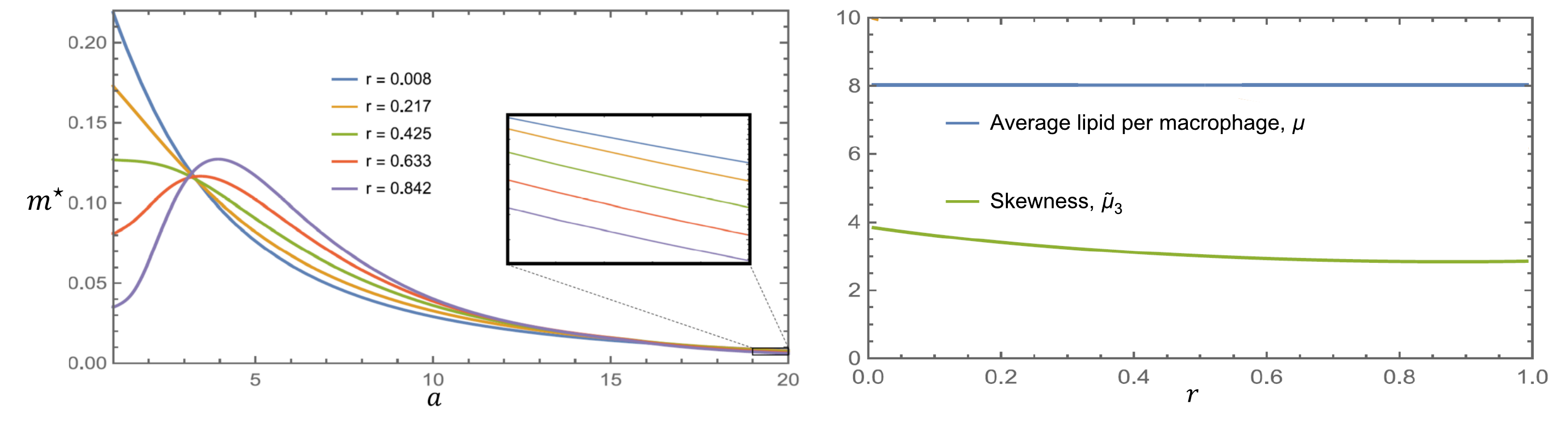}
    \caption{Changes in the relative contribution of proliferation and recruitment affect $m^\star(a)$. The left plot shows $m^\star(a)$ for various values of $r$. We note that the curves do not all pass through a single point upon closer inspection. The right plot shows the mean (which does not change) and skewness as $r$ is increased from 0 to 1. Changes in $r$ were obtained by starting with the parameter set: $\rho = 1/120$, $\gamma = 0.2$, $\eta = 2$, $\theta = 1$, and then raising $\rho$ in increments of $1/120$. The remaining dimensionless parameters were adjusted so that equation \eqref{eqn: alpha rho constraint} holds for all parameter sets used. The cases corresponding to $\rho = 1/120$, $26/120$, $51/120$, $76/120$ and $101/120$ are used for the left plot.}
    \label{fig:m(a) prolif vs recruit}
\end{figure}

By contrast, the equilibrium lipid distribution, $m^\star(a)$, depends sensitively on the relative contribution of proliferation and recruitment. Consider Figure \ref{fig:m(a) prolif vs recruit}, in which we present plots of $m^\star(a)$ subject to the constraints outlined in the above paragraphs. The plots are labelled using the quantity, $r$, which is the proportion of macrophages that are sourced from proliferation, as opposed to recruitment. A simple expression for $r$ can be found by taking the ratio of the total proliferation rate to the combined rate of proliferation and recruitment:
\begin{align}
    r = \frac{\rho M^\star}{\rho M^\star + (A_M^\star - M^\star)/(\kappa + A_M^\star - M^\star)} = \frac{\rho}{1 + \gamma} < 1. \label{eqn: r}
\end{align}
The substantial simplification shown in equation \eqref{eqn: r} is found by substituting the result for $A_M^\star$ in equations \eqref{eqn: ODE steady}. We note that although $m^\star(a)$ changes with $r$, the alterations are anchored by the constraint that the average lipid per macrophage, $A_M^\star/M^\star$, is fixed. The changes therefore manifest in the higher order moments, affecting the skewness.

\section{Discussion} \label{sec: discussion}

% In this paper we extend the lipid-structured model for atherosclerotic plaque development of \cite{ford2019lipid} to account for macrophage proliferation. Although the process is now appreciated as a key event in atherosclerosis \cite{robbins2013local}, the role of macrophage proliferation in plaque development is not well understood. Addressing this gap in knowledge is the focus of our analysis. 

In this paper we present a differential equation population model for macrophage lipid accumulation and necrotic core formation in an atherosclerotic plaque. The defining feature of this model is the use of a structural variable $a$ which represents the lipid contained in macrophages and apoptotic cells, including the endogenous lipid in cell membranes.  From this structured PDE model we are also able to derive an ODE model for the total population of macrophages and apoptotic cells and the lipid that those population contain.

Our model here is novel in that it represents macrophage proliferation in the plaque in addition to efferocytosis, apoptosis and emigration. Although macrophage proliferation is now appreciated as a key event in atherosclerosis \cite{robbins2013local}, the role of the process in plaque development is not well understood. Proliferation introduces nonlocal terms into the partial integro-differential equations first formulated in \cite{ford2019lipid}. Our model equations therefore have commonalities with pantograph-type equations used to model cell proliferation in other contexts (see, for example \cite{hall1989functional,efendiev2018functional}). The model in this paper is more complicated than models with cell populations only, as we are representing, not only cells, but also the lipid loads that they carry. Nevertheless we show that a steady state solution for the PDE can be derived analytically in the special case where there is no efferocytosis (Appendix \ref{secA2}).

Numerical solutions indicate that the model plaque transitions through multiple distinct stages prior to reaching a steady state. These are discussed in Subsection \ref{sec: time-dep}.  Interestingly, the results suggest that much of the necrotic core growth occurs while the number of live and apoptotic macrophage populations are rising very slowly (see $10 \leq t \leq 40$ in Figure \ref{fig:ODE time dependent}). Hence the number of live macrophages may be a poor indicator of early plaque progression when not considered simultaneously with information about plaque lipid content. These observations are also consistent with the work of Lui and Myerscough \cite{lui2021modelling}, and likely manifest due to preferential uptake of apoptotic lipid over necrotic lipid by live macrophages (the assumption $\theta < \eta$ in our model). 

% The growth in necrotic core size is quelled following a rapid increase in live macrophage recruitment at approximately $t = 60$ macrophage lifetimes, which is where necrotic lipid uptake becomes comparable to uptake of apoptotic lipid. The system then gradually tends to steady state. 

The fate of the model plaque and the lipid profile of the population of macrophages in the plaque is determined by the parameters $\rho$ (rate of macrophage proliferation relative to apoptosis), $\gamma$ (rate of macrophage emigration relative to apoptosis), $\eta$ (dimensionless efferocytosis rate) and $\theta$ (dimensionless uptake rate of necrotic lipid). We find that the model tends to a steady state provided that $\rho < 1 + \gamma$, typically doing so within $t = 200$ macrophage lifetimes. The case $\rho \geq 1 + \gamma$ corresponds to the unphysical scenario in which macrophages proliferate faster than they leave the system via apoptosis or emigration, and is not considered in our analysis. The steady state is unique when it exists and numerical solutions indicate that it is stable.

The ODE steady state results show that increases in the proliferation rate produce increases in the number of live macrophages at steady state, $M^\star$, and decreases in eventual necrotic core size, $N^\star$. The result that $M^\star$ increases with $\rho$ is unsurprising and consistent with the experimental results of \cite{tang2015inhibiting} which show that a reduction in macrophage proliferation in mice does, indeed, reduce the number of plaque macrophages. The reduction in $N^\star$
predicted by the model is primarily due to the heightened overall uptake rate of necrotic lipid in the plaque, which is proportional to $\theta M^\star$. However when the efferocytosis rate is sufficiently high ($\eta > \frac{(1+\gamma) \nu}{\lambda}$), increases in the proliferation rate also produce decreases in the amount of apoptotic lipid in the plaque. Hence the decrease in necrotic core size in this regime is also partially attributable to the rapid ingestion of apoptotic cells, and the corresponding decrease in secondary necrosis occurring within the plaque. 

The effects of proliferation are modulated by changes in the other parameters of the model. We find that the efferocytosis parameter $\eta$ is particularly impactful. When $\eta$ is low, increasing the rate of proliferation makes very little difference to the necrotic core size until proliferation rates are close to the upper limit $1 + \gamma$; a large number of plaque macrophages produces no benefit if they are ineffective at efferocytosis. Poor efferocytosis is known to be a feature of vulnerable plaques \cite{tabas2010macrophage} and this modelling suggests that increasing proliferation is unlikely to be a good therapeutic intervention unless macrophages are also effectively removing apoptotic cells and necrotic material. If efferocytosis is defective, proliferation merely introduces more lipid into the plaque macrophages and will therefore increase the lipid in apoptotic  cells and thence, via secondary necrosis, the amount of lipid in the necrotic core.

Emigration, which increases with $\gamma$, appears to have a mixed effect on necrotic core size. For proliferation rates $\rho < 1$, increases in $\gamma$ decrease core size until $\gamma \approx 0.15$ and increase core size thereafter. This trend likely occurs because plaques with little emigration contain macrophages with a higher lipid burden. When these lipid-laden macrophages emigrate from the lesion their internalised lipid is also removed from the system and so does not contribute to the necrotic core. If instead emigration is frequent, macrophages leave the plaque before they can phagocytose significant amounts of necrotic lipid, and so the core grows due to lack of necrotic lipid uptake. For proliferation rates $\rho \geq 1$, the results indicate that increases in $\gamma$ monotonically increase $N^\star$. Hence the model predicts that increasing the migratory propensity of a self-replenishing macrophage population will always increase necrotic core size. 

Steady state analysis of the full model indicates that proliferation contributes to an overall decrease in macrophage lipid burden. The resulting macrophage lipid distribution, $m^\star (a)$, becomes increasingly peaked at lower values of $a$ with a correspondingly thinner tail as $\rho$ is increased. We note that the global mode occurs at some $a > 1$ for large enough proliferation rates, in contrast to the profiles observed in \cite{ford2019lipid}. Given that proliferation is lower in early stage plaque than in late stage plaque in mice, \cite{robbins2013local}, if the model is correct, then we would expect the ratio of relatively unladen macrophages (say $1 < a < 3$) to heavily laden macrophages (say $a > 10$) to be higher in late stage plaques than in early plaques. 

The results of the model presented here, as with all conceptual models, need to be interpreted carefully and with the model assumptions in mind.  In particular, this model is built on the assumption that macrophages' behaviours are not dependent on their lipid load.  Observations suggest that emigration and proliferation, for example, both depend on a cell's lipid load \cite{pataki1992endocytosis, kim2018transcriptome}. Lipid-dependent macrophage behaviour is the focus of the study Watson \textit{et al.} (2022, in preparation).

This model, notably, does not contain any mechanism for resolving plaque growth or for plaque regression.  To include such behaviour in the model requires either carefully designed functions for lipid dependent behaviour or the introduction of other terms or other species (such as M2 macrophages) into the model \cite{Tabas_Bornfeldt_2016}.

In conclusion, we present here a model for populations of live macrophages and apoptotic macrophages in an atherosclerotic plaque where the populations are structured by the amount of lipid each cell contains and where live macrophages can proliferate.  We show that in the model, macrophage proliferation generally increases the number of macrophages and increases the proportion of plaque lipids that are inside macrophages, rather than in the necrotic core.  In plaques where efferocytosis is effective, macrophage proliferation can significantly reduce the size of the necrotic core.  In a proliferative population of plaque macrophages there will be fewer macrophages with very high lipid loads, with macrophages predominantly attaining a small but non-zero lipid burden. By splitting lipid loads when foam cells divide, proliferation spreads the macrophage lipid burden across the population.

\backmatter

%\bmhead{Supplementary information}

% If your article has accompanying supplementary file/s please state so here. 

% Authors reporting data from electrophoretic gels and blots should supply the full unprocessed scans for key as part of their Supplementary information. This may be requested by the editorial team/s if it is missing.

% Please refer to Journal-level guidance for any specific requirements.

\bmhead{Acknowledgments}
We acknowledge funding (to MRM) from the Australia Research Council Discovery Program, grant number DP200102071.
% Acknowledgments are not compulsory. Where included they should be brief. Grant or contribution numbers may be acknowledged.

% Please refer to Journal-level guidance for any specific requirements.

%\section*{Declarations}

% Some journals require declarations to be submitted in a standardised format. Please check the Instructions for Authors of the journal to which you are submitting to see if you need to complete this section. If yes, your manuscript must contain the following sections under the heading `Declarations':

% \begin{itemize}
% \item Conflict of interest/Competing interests (check journal-specific guidelines for which heading to use)
% \item Ethics approval 
% \item Consent to participate
% \item Consent for publication
% \item Availability of data and materials
% \item Authors' contributions
% \end{itemize}

% \noindent
% If any of the sections are not relevant to your manuscript, please include the heading and write `Not applicable' for that section. 

% %%===================================================%%
% %% For presentation purpose, we have included        %%
% %% \bigskip command. please ignore this.             %%
% %%===================================================%%
% \bigskip

% Springer journals and proceedings: \url{https://www.springer.com/gp/editorial-policies}

% Nature Portfolio journals: \url{https://www.nature.com/nature-research/editorial-policies}

% \textit{Scientific Reports}: \url{https://www.nature.com/srep/journal-policies/editorial-policies}

% BMC journals: \url{https://www.biomedcentral.com/getpublished/editorial-policies}
% \end{flushleft}

\newpage
\begin{appendices}

\section{ODE equilibrium trends}\label{secA1}
Here we provide a proof of the trends mentioned in Subsection \ref{sec: ODE parameter dependence} by explicitly considering partial derivatives with respect to the proliferation parameter, $\rho$.

\begin{itemize}
    \item $M^\star$: \\
We note that the solution \eqref{eqn: Mstar} can be written as:
\begin{align}
    M^\star = \frac{-k_1 + \sqrt{k_1^2 + 4k_2}}{2}, 
\end{align}
where 
\begin{align}
    k_1 &= \frac{\lambda - 1 + \gamma (\kappa + \gamma \kappa + \lambda) - \rho (\lambda + \gamma \kappa)}{1 + \gamma - \rho}, & k_2 &= \frac{\lambda}{1 + \gamma - \rho}.
\end{align}
It follows from the chain rule that:
\begin{align}
    \frac{\partial M^\star}{\partial \rho} &= \frac{\partial M^\star}{\partial k_1} \frac{\partial k_1}{\partial \rho} + \frac{\partial M^\star}{ \partial k_2} \frac{\partial k_2}{\partial \rho}, \nonumber \\
    &= \frac{1}{2 (1 + \gamma - \rho)^2} \Big( \frac{k_1}{\sqrt{k_1^2 + 4k_2}} - 1 \Big) + \frac{\lambda}{(1 + \gamma - \rho)^2} \frac{1}{\sqrt{k1^2 + 4k_2}}. \label{eqn: M sensitivity}
\end{align}
Since $k_2 > 0$ (using that $\rho < 1 + \gamma$ as noted in Section \ref{sec: parameter treatment}), the first term is positive and hence also $\frac{\partial M^\star}{\partial \rho} > 0$. \\

    \item $P^\star$: \\
From the relation for $P^\star$ in \eqref{eqn: ODE steady}, we note that $P^\star$ is an increasing function of $M^\star$. From the result proved above, we also have $\frac{\partial P^\star}{\partial \rho}$. \\

    \item $A_M^\star$: \\
From the relation for $A_M^\star$ in \eqref{eqn: ODE steady}, we can write:
\begin{align}
    A_M^\star - M^\star = \kappa \Big( \frac{1}{1+ (\rho - 1 - \gamma) M^\star} - 1 \Big).
\end{align}
Differentiating the above line with respect to $\rho$ gives:
\begin{align}
    \frac{\partial A_M^\star}{\partial \rho} - \frac{\partial M^\star}{\partial \rho} &= \kappa \frac{1 + \gamma - \rho}{(1 + M^\star [1 + \gamma - \rho]^2)} \frac{\partial M^\star}{\partial \rho}.
\end{align}
Using that $\frac{\partial M^\star}{\partial \rho} > 0$, we find that $\frac{\partial A_M^\star}{\partial \rho} > \frac{\partial M^\star}{\partial \rho} > 0$. \\

    \item $N^\star$: \\
The relation for $N^\star$ in \eqref{eqn: ODE steady} can be written as:
\begin{align}
    N^\star = \frac{\nu}{\theta} \frac{A_P^\star}{A_M^\star} \cdot \frac{A_M^\star}{M^\star}.
\end{align}
The fraction $\frac{A_P^\star}{A_M^\star}$ is a decreasing function of $M^\star$, using the relation for $A_P^\star$ in \eqref{eqn: ODE steady}. Hence $\frac{A_P^\star}{A_M^\star}$ is also a decreasing function of $\rho$, using that $\frac{\partial M^\star}{\partial \rho} > 0$. The fraction $\frac{A_M^\star}{M^\star}$ can be written as
\begin{align}
    \frac{A_M^\star}{M^\star} = 1 + \frac{\kappa}{\frac{1}{1 + \gamma - \rho} - M^\star}.
\end{align}
The denominator, $\frac{1}{1 + \gamma - \rho} - M^\star$, can be shown to be an increasing function of $\rho$ using \eqref{eqn: M sensitivity}. Hence, $\frac{\partial}{\partial \rho} (A_M^\star / M^\star) < 0$. It then follows that $\frac{\partial N^\star}{\partial \rho} < 0 $ since $N^\star$ is the product of two positive and decreasing functions of $\rho$. \\

\item $A_P^\star$: \\
By differentiating the expression for $A_P^\star$ in \eqref{eqn: ODE steady} using the chain rule, we can write:
\begin{align}
    \frac{\partial A_P^\star}{ \partial \rho} &= \frac{\partial A_P^\star}{\partial A_M^\star} \frac{\partial A_M^\star}{\partial \rho} + \frac{\partial A_P^\star}{\partial M^\star} \frac{\partial M^\star}{\partial \rho}, \nonumber \\
    &= \frac{1}{\eta M^\star + \nu} \Big( \frac{\partial A_M^\star}{\partial \rho} - \eta A_P^\star \frac{\partial M^\star}{\partial \rho} \Big).
\end{align}
By analysing the sign of the bracketed term, we find that $\frac{\partial A_P^\star}{\partial \rho} > 0$ when $\eta < \frac{(1 + \gamma) \nu}{\lambda}$ and $\frac{\partial A_P^\star}{\partial \rho} < 0$ if $\eta > \frac{(1 + \gamma) \nu}{\lambda}$. The singular case, $\eta = \frac{(1+ \gamma) \nu}{\lambda}$ gives $\frac{\partial A_P^\star}{\partial \rho} = 0$.

\end{itemize}

\newpage
\section{Analytical solution for $m^\star (a)$ when $k_1 = 0$}\label{secA2}

Taking the Laplace transform of equation \eqref{eqn: mstar ode} gives the functional algebraic relation:
\begin{align}
    s \hat{m}(s) - m^\star_1 = k_1 e^{-s} \hat{m}(s)^2 - k_2 \hat{m}(s) + \frac{k_3}{2} \hat{m}(s/2), \label{eqn: laplace}
\end{align}
where
\begin{align}
    \hat{m}(s) := \int_0^\infty m^\star (a+1) e^{-as} da.
\end{align}
Solving equation \eqref{eqn: laplace} when $k_1 = 0$ gives:
\begin{align}
    \hat{m}(s) =  \frac{m^\star_1}{s + k_2} + \frac{k_3}{2 (s + k_2)} \hat{m}(s/2). \label{eqn: m(s) functional}
\end{align}
Equation \eqref{eqn: m(s) functional} can be solved by iteration. Assuming that Re$(s) > 0$, recursive application of equation \eqref{eqn: m(s) functional} gives
\begin{align}
    \hat{m}(s) &= \frac{m^\star_1}{s + k_2} + \frac{k_3}{2 (s + k_2)} \hat{m}(s/2) \nonumber \\
    &= \frac{m^\star_1}{s + k_2} + \frac{k_3}{2 (s + k_2)} \Big( \frac{m^\star_1}{s/2 + k_2} + \frac{k_3}{2 (s/2 + k_2)} \hat{m}(s/4) \Big) \nonumber \\
    &\, \, \, \vdots \nonumber \\
    &= m^\star_1 \sum_{j = 0}^n \frac{(k_3/2)^j}{\prod_{\ell=0}^j (s/2^\ell + k_2)} + \frac{(k_3/2k_2)^{n+1}}{\prod_{j=0}^n (1 + k_2 s/2^j)} \hat{m}(s/2^{n+1}),
\end{align}
for every $n \geq 0$. We note that the second term tends to zero as $n \rightarrow \infty$. To see this, recall first that $\hat{m}(0) = 1$ from the non-dimensionalisation \eqref{eqn: nondim}. Substituting $s = 0$ into equation \eqref{eqn: m(s) functional} gives $k_3/(2k_2) = 1 - m^\star_1/k_2 < 1$. It then follows from our assumption that Re$(s) > 0$ and the continuity of $\hat{m}$ at $s = 0$ that
\begin{align}
    \Bigg\vert \frac{(k_3/2k_2)^{n+1}}{\prod_{j=0}^n (1 + k_2 s/2^j)} \hat{m}(s/2^{n+1}) \Bigg\vert \leq \Big( \frac{k_3}{2k_2} \Big)^{n+1} \vert \hat{m}(s/2^{n+1}) \vert \rightarrow 0
\end{align}
as $n \rightarrow \infty$. Hence,
\begin{align}
    \hat{m}(s) =  m^\star_1 \sum_{j = 0}^n \frac{(k_3/2)^j}{\prod_{\ell=0}^j (s/2^\ell + k_2)}. \label{eqn: m(s) sol}
\end{align}
The product in equation \eqref{eqn: m(s) sol} can be decomposed into a sum of partial fractions. Doing so allows us to write
\begin{align}
    \hat{m}(s) = \frac{m^\star_1}{k_2} \sum_{j=0}^\infty  \Big( \frac{k_3}{2k_2} \Big)^j  \sum_{i=0}^j \frac{1}{\prod_{\ell = 0, \ell \neq i}^j (1-2^{i-\ell})} \frac{1}{1 + k_2 s/2^i}.
\end{align}
Finally, we take the inverse Laplace transform term-by-term (justified by the uniform convergence of the above series) to obtain the solution \eqref{eqn: no use}.
% An appendix contains supplementary information that is not an essential part of the text itself but which may be helpful in providing a more comprehensive understanding of the research problem or it is information that is too cumbersome to be included in the body of the paper.




\end{appendices}


\bibliography{references}% common bib file

% \input sn-sample-bib.tex%

\end{document}